\documentclass[a4paper, 12pt]{article}

\usepackage{fancyhdr}
\usepackage{amsmath}
\usepackage{graphicx}
\usepackage{url}
\usepackage{amsbsy,amssymb} 
\usepackage{amsmath}
\usepackage{hyperref}
\usepackage{color}
\usepackage{authblk}

\usepackage{fancyvrb}

\usepackage{rotating}

\usepackage[caption = false]{subfig}

\usepackage{lscape}

\title{A framework for assumption-free assessment of imperfect geometry of a linac C-arms}
\author[1,2]{R. A. Kycia}
\author[1]{Z. Tabor}
\author[1]{A. Woszczyna}
\author[3]{D. Kabat}
\author[1]{Z. Lata\l a}
\author[3]{M. Tulik}
\affil[1]{Cracow University of Technology, Faculty of Physics, Mathematics and Computer Science, PL-31155, Krak\'ow, Poland}
\affil[2]{Department of Mathematics and Statistics, Masaryk University, Brno, Czechia}
\affil[3]{Centre of Oncology, Maria Sklodowska-Curie Memorial Institute, Krak\'ow, Poland}

\begin{document}
\maketitle
\begin{abstract}
\noindent
In the present paper a general setup for determination of imperfect geometry of radiotherapeutic devices has been proposed that base on geometric algebra framework. To account for this imperfect geometry, two methods of a calibration were presented, consisting of determining for each angular position of a gantry a correction shift which must be applied to the origin of a laboratory frame of reference to place it along a radiation axis for this angular position. Closed form solutions for these corrections are provided.
\end{abstract}


\textbf{Running title}: Geometric algebra for cancer treatment

\textbf{Classification}: 51A05, 51P05, 70Q05

\section{ Introduction }
Geometric algebra (GA) as a part of Clifford algebras is an efficient tool for performing vector manipulations, rotations and projections. In the past, it was competing with well known mixed approach to such computations known as vector algebra promoted by Gibbs \cite{VectorWars} and lost. Therefore currently we use the objects which results from cross product (e.g., angular momentum, angular velocity, torque) and call them 'vectors', which they are not and in fact they describe higher graded objects (here planes) that naturally occur in geometric algebra. Recently, due to the revival of the Clifford/Hilbert/Grassmann ideas by David Hastenes and co workers \cite{Hastens_DesignOfAlgebra}, \cite{GeomALgForPhysicists} it returns to the standard curriculum of scientists and engineers.

Currently, one of the most important areas of application of projective geometry in the field of life sciences is external beam therapy (EBT) which is the most common method used in the treatment of cancer deseases. During EBT a source of ionizing radiation (which is a part of a linac - a linear medical accelerator) moves along a trajectory which should be circular in an ideal case. While moving, the source is emitting a beam of high energy photons. The photons deposit energy in the tissues of a patient. The goal of the therapy is to design a therapy plan in such a way that the photon energy is deposited primarily in the cancer tissues while healthy tissues remain intact. 

Another important task directly related to EBT is periodic exploitation testing of medical equipment used in EBT. For example, due to inaccuracies in the construction and weights of linac components, the actual position of each of the moving elements of an accelerator may differ slightly from the planned ones. Due to flexing or sagging of a linac components under their own weight the trajectories of these components differ from ideal. To keep these inaccuracies under control periodic assessement of the geometry of EBT devices must be conducted which include among others evaluation of an isocenter position, a precision of a gantry, a collimator and a couch movements \cite{Geom1}, \cite{Geom2}. Whenever observed inaccuracied exceed a tolerance limit an intervention of a technical service becomes necessary.

Both EBT planning and exploitation testing of EBT devices rely havily on the projective geometry. Certainly, among the exploitation tests the most important one is the determination of the isocenter position \cite{Isocenter} which is a region of 3D space (ideally a point) where a cancer should be positioned in order to provide the most effective treatment. Determination of the isocenter involves examining projective images of special phantoms usually containing multiple fiducial ball bearings (BBs) \cite{Phantom1} - \cite{Part1} although application of other kinds of fiducial markers is also possible \cite{Part1}. 

To assess the geometry of an EBT device the spatial relationships between the fiducial markers and the components of a therapeutic device must be analyzed. Published solutions are essentially based on formulas which are derived ad hoc for an assumed phantom design and device geometry. In contrast, in the present paper we will use solely GA for developing an universal and the most general assumption-free approach to the calibration of EBT devices, where by calibration we mean determination of various geometric aspects of EBT devices. 

Th paper is organized as follows: In the next sections we formulate the problem and then we provide short introduction to Geometric Algebra. Next general discussion of the projection will be presented. We also present measurement technique and finally the calibration algorithm. In the appendix we present an introduction to help in recovering our results in Mathematica CAS using 'Cartan' package \cite{GAMathematica} with some additional code which shows how to perform some lengthy derivations which were omitted in main text. Precise equations for cubic phantom vertices projections will be also provided and GA-based method for radiotherapeutic device calibration will be discussed.

The original contribution of the present paper is development of GA framework for comprehensive assessment of geometry of EBT devices.

\section{Problem formulation}

The present paper is a continuation of our previous work \cite{Part1}, where one of possible phantom designs was proposed together with an optimization least-square method for assessment of various geometric characteristics (like isocenters) of EBT devices. We presented a method that can determine the geometry of a device using at least $13$ parameters which, given three additional constraints, requires measuring the position in the detector plane of projections of at least $5+1$ fiducial points in 3D space, the coordinates of which are exactly known ($2$ variables per point). This minimal number is required for the method to work, however, as optimization is used, therefore the more fiducial markers are used the better estimation is. The outline of the method is as follows: place the global coordinate system with orthonormal base $\{e_{1},e_{2},e_{3}\}$ at the center $O$ of the phantom made of some number of fiducial elements (e.g. balls). Then the source of ionizing radiation is determined by the source position $Z=(Z_{1}, Z_{2},Z_{3})$. The projection (or equivalently detector) plane is determined by a unit vector $n$, which is perpendicular to the detector plane and the source to detector distance $L$. On the projection plane we have no hint on the orientation of the projected base of the global frame, therefore, we use an arbitrary frame with the axes:
\begin{equation}
 \begin{array}{c}
  E_{1}=\sum_{i=1}^{3}\lambda^{i}_{1}e_{1}, \\
  E_{2}=\sum_{i=1}^{3}\lambda^{i}_{2}e_{2},
 \end{array}
\end{equation}
This arbitrarily introduces six new parameters $\lambda^{i}_{j}$, which have to be determined. The task in hand is to reconstruct various geometric characteristics of an EBT device, including source position, normal vector $n$, distance $L$, and coefficients $\lambda^{i}_{j}$ given only projection images of fiducial markers in the projection (detector) plane. Additionally, during therapy the source (enclosed within so called gantry) moves along some trajectory which has to be reconstructed too.

 The proposed GA-based calibration procedure returns for each nominal angular position of a gantry a 3D correction vector and fully specified trajectory of a source and a detector plane. The correction vector provides information about how an origin of a laboratory frame of reference must be optimally shifted to cross the axis of a radiation field (the axis which should cross a cancer during an actual therapy).

We use Clifford algebra/geometric algebra approach \cite{Hastens_DesignOfAlgebra}, \cite{Hastens_Projective}, \cite{GeomALgForPhysicists} as it provides an elegant and consistent framework for the problem in hand. An introduction to GA will be presented in the next section. Note that there are simple packages for effective GA manipulation, such as \cite{GAMathematica} for Mathematica Computer Algebra System(CAS).

\section{Overview of Geometric Algebra}

\subsection{Short introduction to Geometric Algebra}
This short introduction is based on \cite{Hastens_DesignOfAlgebra}, \cite{GeomALgForPhysicists} and we refer interested reader to these sources for further references.

Assume that we have basis vectors $\{e_{1}, e_{2}, e_{3}\}$, where inner product $e_{i}\cdot e_{i}=1$ and $e_{i}\cdot e_{j} = 0$ where $i\neq j$. For any two vectors $a=a_{1}e_{2}+a_{2}e_{2}+a_{3}e_{3}$ and $b=b_{1}e_{2}+b_{2}e_{2}+b_{3}e_{3}$ we know from basic algebra the inner product:
\begin{equation}
 a \cdot b = a_{1}b_{1} + a_{2}b_{2}+a_{3}b_{3}.
\end{equation}
The result is a scalar.

We now introduce new operation on vectors the outer product, which is antisymmetric operation
\begin{equation}
 a \wedge b = - b \wedge a,
\end{equation}
from which results that 
\begin{equation}
 a \wedge a = 0.
\end{equation}
It defines a new object which is, so called, higher grade object - it is not a scalar or vector. In the case of two non-parallel vectors it is an oriented plane, called a bivector(or a blade or sometimes called quaternions) which is spanned by these vectors oriented from the first vector to the second one to the product.

Finally, we introduce another kind of product which mixes above-defined two products, called geometric product. It is denoted using no sign between terms as
\begin{equation}
 ab = a\cdot b + a \wedge b.
\end{equation}
The product is a sum of a different grade components - a scalar one and a bivector. It is useful to express this product on the base vectors
\begin{equation}
 \begin{array}{c}
  e_{i}e_{j} = e_{i} \wedge e_{j} = -e_{j} \wedge e_{i} = -e_{j}e_{i}, \quad i \neq j; \\
  e_{i}e_{i} = 1.
 \end{array}
\end{equation}

Geometric algebra is a framework which is ideally suitable for the description of rotations and this prescription we will need in what follows. The object useful in rotation is a rotor object
\begin{equation}
 R(\theta) = e^{-B\frac{\theta}{2}}, \quad B^2 = -1,
\end{equation}
where $B$ is a bivector defining an oriented plane in which rotation is performed and $\theta$ is the rotation angle. Expanding the rotor using Taylor's series for exponent we obtain
\begin{equation}
 R(\theta) = \cos(\theta / 2) - B \sin(\theta / 2 ).
\end{equation}

We also define a reversed rotor $R^{\dagger}$ in which we reverse order of all vectors in products. Now the rotation of a vector $a$ can be described by the famous double-group prescription
\begin{equation}
 a' = RaR^{\dagger}.
\end{equation}

Usually, the rotations are performed in the planes described by the base vectors, e.g., rotations defined by the Euler's angles. As an example, we present a rotation in the $e_{1}\wedge e_{2}$ plane.
In order to rotate $e_{1}$, $e_{2}$ and $e_{3}$ first we check the commutativity with the blade
\begin{equation}
\begin{array}{c}
 e_{1} e_{1}e_{2} = -e_{1}e_{2}e_{1}, \\
 e_{2} e_{1}e_{2} = -e_{1}e_{2}e_{2}, \\
 e_{3} e_{1}e_{2} = -e_{1} e_{3}e_{2} = e_{1}e_{2}e_{3}.
\end{array}
\end{equation}
These rules can be transported to rotors, which gives
\begin{equation}
 \begin{array}{c}
  e^{-e_{1}e_{2}\frac{\theta}{2}}e_{1}e^{e_{1}e_{2}\frac{\theta}{2}}=e_{1}e^{e_{1}e_{2}\theta}= e_{1}(\cos(\theta)+e_{1}e_{2}\sin(\theta)) = e_{1}\cos(\theta)+e_{2}\sin(\theta), \\
  e^{-e_{1}e_{2}\frac{\theta}{2}}e_{1}e^{e_{1}e_{2}\frac{\theta}{2}} = e_{2}( cos(\theta) + e_{1}e_{2}\sin(\theta)) = -e_{1}\sin(\theta) + e_{2}\cos(\theta), \\
  e^{-e_{1}e_{2}\frac{\theta}{2}}e_{3}e^{e_{1}e_{2}\frac{\theta}{2}} = e_{3}e^{0} = e_{3},
 \end{array}
\end{equation}
as it can be done be obtained using rotation matrix.

The prescription of rotating the unit vector $a$ onto the unit vector $b$ in the plane spanned by these vectors can be calculated using the well-known fact that rotation can be decomposed into two reflections (see \cite{GeomALgForPhysicists}, section 2.7.2), and we get
\begin{equation}
 R=\frac{1+ab}{\sqrt{2(1+a\cdot b)}} = e^{\frac{b\wedge a}{\sin(\theta)} \frac{\theta}{2}},
 \label{Eq.RotorFromAtoB}
\end{equation}
where $\theta$ is the angle between the vectors $a$ and $b$.

We can define the dualization operation in three dimensions, which uses $I=e_{1}e_{2}e_{3}$, which in three dimensions helps to transform a bivector $a\wedge b$ into grade $1$ object ('a vector') associated to the bivector and explains what the well know cross product is
\begin{equation}
 a\times b = -I(a\wedge b).
 \label{Eq.dualization}
\end{equation}
The formula explains the true nature of the cross product and its unusual behaviour. It is obvious that the cross product is possible only in three dimensions, however wedge product has no such limitation and describe the same type of geometric notion.

\subsection{Projective geometry using geometric algebra}
For the detailed introduction to the projective geometry consult \cite{Hastens_Projective} or \cite{GeomALgForPhysicists}.

\begin{figure}
\centering
 \includegraphics[width = 0.5\textwidth]{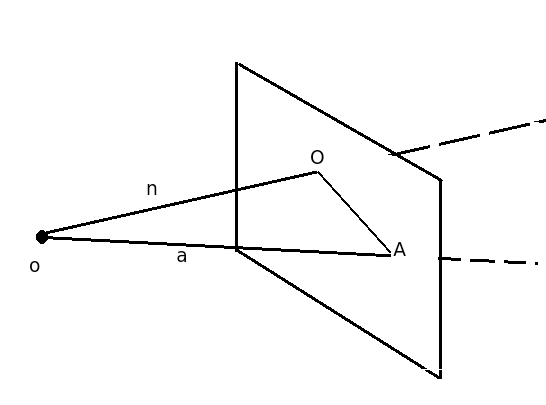}
 \caption{Projective plane.}
 \label{Fig.ProjectivePlane}
\end{figure}

Assume that we have a point $o$, which is projected perpendicularly along $n$ vector to the plane. Its image is $O$ point. Consider now the vector $a$. The task is to find the projection vector $OA$, see Fig. \ref{Fig.ProjectivePlane}.

We have $n+OA=\lambda a$, where $\lambda$ is a constant which we want to eliminate. Multiplying by $n$ one gets $\lambda = n^{2}(a\cdot n)^{-1}$ and then 
\begin{equation}
 OA = \frac{a \wedge n}{a \cdot n} n.
 \label{projectiveSplit}
\end{equation}
This formula defines a projective split which will be extremely useful in our further considerations. Only $n$ and some vector $a$ from the line $\lambda a$ have to be known in order to define a projection. The formula (\ref{projectiveSplit}) can be simplified using $a \wedge n = an - a \cdot n$, which gives
\begin{equation}
 a = \frac{an^{2}-n(a \cdot n)}{a \cdot n}.
\end{equation}

In the next sections we will use this information to define the algorithms for assessment of thegeometry of an EBT device.

\section{General considerations}

\begin{figure}
\centering
 \includegraphics[width = 0.5\textwidth]{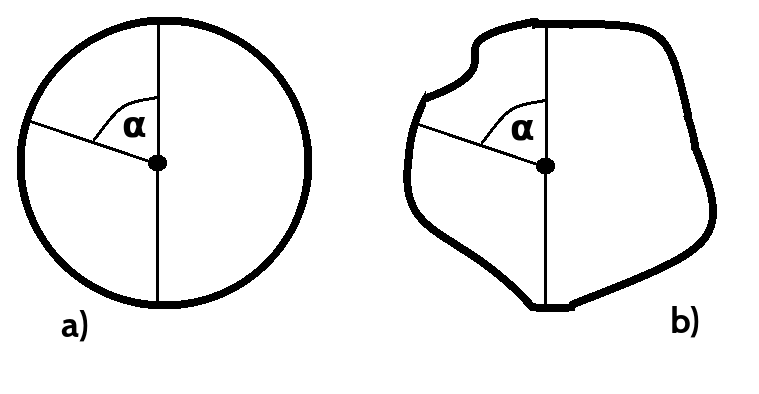}
 \caption{Gantry trajectory and $\alpha$ parameter as an angle from some fixed direction. a) represents ideal situation which is deviated be elasticity of material and gravitational force, which is presented in an exaggerated way in b).}
 \label{Fig.GantryTrajectory}
\end{figure}

We assume that a gantry of an EBT device moves along a continuous curve around some 'center' and this curve can be uniquely parametrized by some parameter $\alpha$ (for example the nominal angle of rotation around the center) - see Fig. \ref{Fig.GantryTrajectory}. This parameter is usually displayed at an operator console of a device. We also assume that the parameterization of a gantry trajectory by the value of $\alpha$ is reproducible if we want to rely on calibration results during therapy of any individual patient.

\begin{figure}
\centering
 \includegraphics[width = 0.5\textwidth]{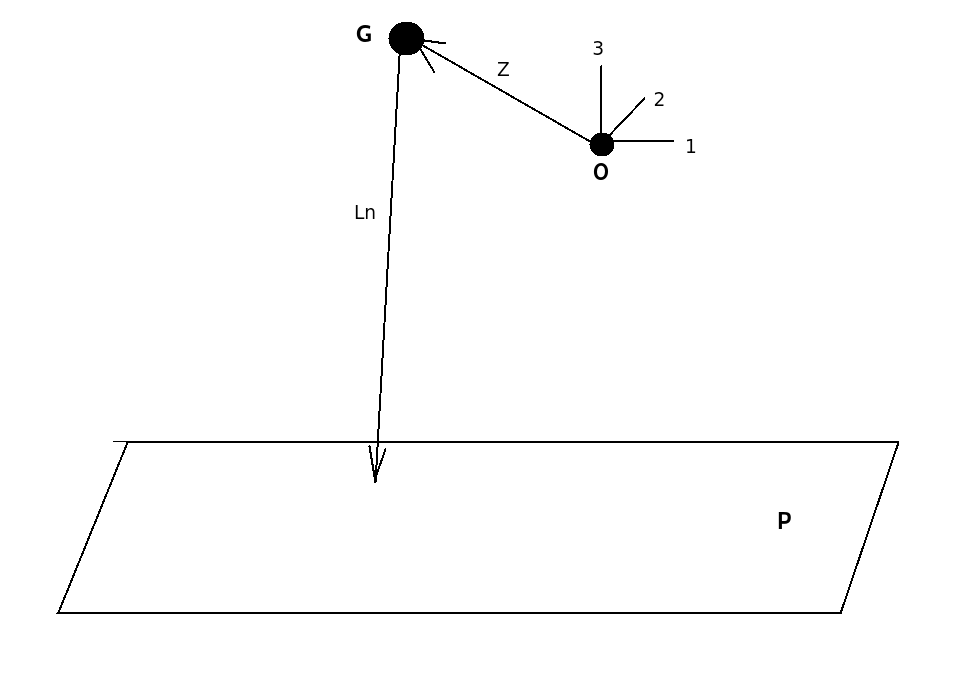}
 \caption{General setup of the calibration system. $0$ denotes the origin of phantom reference frame, $1$ denotes $e_{1}$ vector etc. G - denotes position of the radiation source in the gantry and $Z$ is the vector describing the position of the source in the global reference frame and $L$ is the distance source-imaging plane and $n$ gives the unit vector perpendicular to the imaging plane $P$ and on the line source-plane.}
 \label{Fig.GeneralSetupCalibration}
\end{figure}

The origin of the global reference frame is placed in a center of the phantom used for calibration as it is presented in Fig. \ref{Fig.GeneralSetupCalibration} and the axes of the reference frame are aligned with the phantom (e.g. with its edges if a phantom has a cubical shape). The vector $Z=(Z_{1},Z_{2},Z_{3})$ gives the position to the center $G$ of the radiation source and depends on $\alpha$ as well as the vector $n=(n_{1},n_{2},n_{3})$ which is unit ($n\cdot n =1$), and the distance $L$ of $G$ to the detector plane $P$.

\subsection{Isocenter of radiation field alignment calibration}
The most important during the radiotherapy is the crossing of the beam central line with cancer. For an ideal system a beam central line should thus coincide with $Z$, it is the most important to provide alignment of these two vectors. 

\begin{figure}
\centering
 \includegraphics[width = 0.5\textwidth]{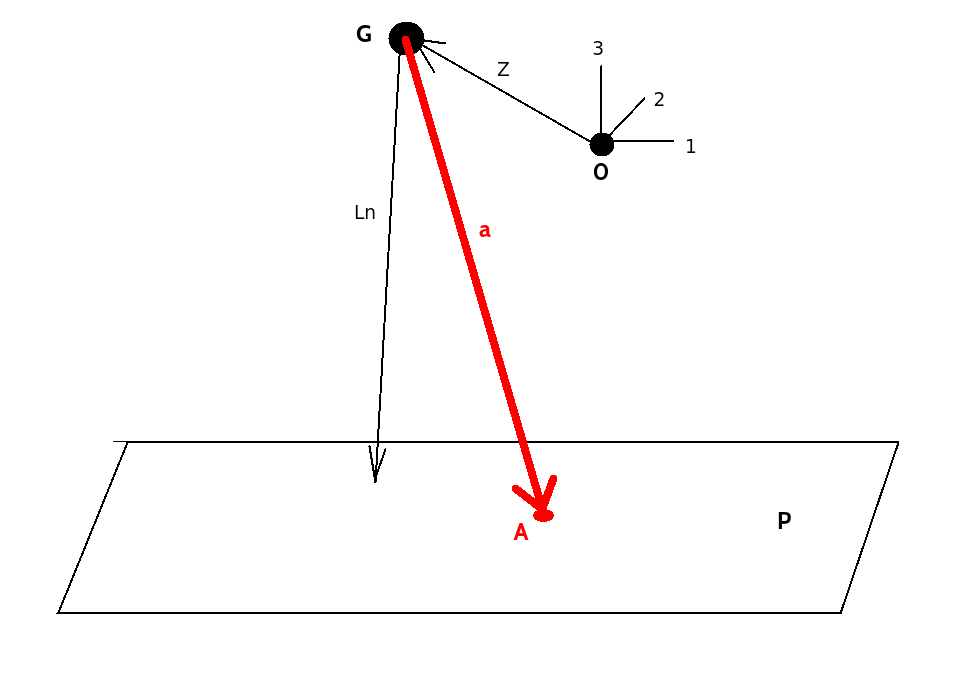}
 \caption{General setup of the calibration system with source center given by vector $a$ and its projection on the plane $P$ at the point $A$.}
 \label{Fig.GeneralSetupCalibrationSource}
\end{figure}

We assume that the central line of the beam has been selected for a given parameter $\alpha$ and can be geometrically described on the projection plane $P$, as presented in Fig. \ref{Fig.GeneralSetupCalibrationSource} by the point $A$.
The vector $a=(a_{1},a_{2},a_{3})$ presents the beam center line direction. 

In the figure the alignment is also not perfect which would be when
\begin{equation}
 a \wedge Z = 0.
 \label{Eg.BeamAlignment}
\end{equation}
This gives the condition
\begin{equation}
 (a_{2}Z_{1}-a_{1}Z_{2})e_{1}e_{2}+(a_{3}Z_{1}-a_{1}Z_{3})e_{1}e_{3}+(a_{3}Z_{2}-a_{2}Z_{3})e_{2}e_{3} =0,
\end{equation}
which is equivalent to vanishing of components of the vector product $a \times Z$ by dualization procedure (\ref{Eq.dualization}).

We can provide correction shift for phantom center point in order to make perfect alignment for given angle $\alpha$, that is to place a phantom center in the central axis of a radiation field:
\begin{equation}
 \Delta Z_{\alpha}=\frac{Z \wedge a}{a^{2}}a,
 \label{Eq.RadiationSourceCorrection}
\end{equation}
which can easily rewritten in the more familiar form
\begin{equation}
 \Delta Z_{\alpha} = \frac{Z a^{2}-(Z\cdot a) a}{a^{2}},
\end{equation}
from which it can be seen that it is the part of $Z$ perpendicular to the vector $a$.

\subsection{Projections}
In this subsection the equations for projections using geometric algebra will be provided. They can be the most effectively derived using computer algebra packages described in the appendix \ref{Appendix:MathematicaPackage}.

The first projection is for the center of the reference frame as seen from the position of the source - it is $-Z=(-Z_{1},-Z_{2},-Z_{3})$ and projects onto the detector plane to:
\begin{equation}
\begin{array}{c}
 E_{0}=\frac{Z \wedge n }{Z \cdot n}Ln =  \\
 \frac{L \left({n_1}^2 \left(e_2 {Z_2}+e_3 {Z_3}\right)-{n_2} \left(e_2
   {n_1} {Z_1}+e_1 {n_1} {Z_2}+e_3 {n_3} {Z_2}+e_2 {n_3}
   {Z_3}\right)-{n_1} {n_3} \left(e_3 {Z_1}+e_1
   {Z_3}\right)+{n_2}^2 \left(e_1 Z_1+e_3 Z_3\right)+{n_3}^2
   \left(e_1 {Z_1}+e_2 {Z_2}\right)\right)}{{n_1}{Z_1}+{n_2}
   {Z_2}+{n_3}{Z_3}}.
\end{array}
\end{equation}

Next, we project the base vectors $e_{1}$, $e_{2}$ and $e_{3}$ onto the plane
\begin{equation}
 \begin{array}{c}
  E_{1} = \frac{(-Z+e_{1})\wedge n}{(-Z+e_{1})\cdot n} Ln - E_{0} = \\
  \frac{L \left({n_1}^2+{n_2}^2+{n_3}^2\right) \left(e_2 {n_1} {Z_2}+e_3
   {n_1} {Z_3}+e_1 (-{n_2}) {Z_2}-e_1 {n_3}
   {Z_3}\right)}{({n_1} ({Z_1}-1)+{n_2} {Z_2}+{n_3} {Z_3})
   ({n_1} {Z_1}+{n_2} {Z_2}+{n_3} {Z_3})}, \\ \\
    E_{2} = \frac{(-Z+e_{2})\wedge n}{(-Z+e_{2})\cdot n} Ln - E_{0} = \\
    -\frac{L \left({n_1}^2+{n_2}^2+{n_3}^2\right) \left(e_2 ({n_1}
   {Z_1}+{n_3} {Z_3})-{n_2} \left(e_1 {Z_1}+e_3
   {Z_3}\right)\right)}{({n_1} {Z_1}+{n_2} ({Z_2}-1)+{n_3}
   {Z_3}) ({n_1} {Z_1}+{n_2} {Z_2}+{n_3} {Z_3})}, \\ \\
   E_{3} = \frac{(-Z+e_{3})\wedge n}{(-Z+e_{3})\cdot n} Ln - E_{0} = \\
    -\frac{L \left({n_1}^2+{n_2}^2+{n_3}^2\right) \left( -n_{3}(Z_{1}e_{1}+Z_{2}e_{2})+(n_{1}Z_{1}+n_{2}Z_{2})e_{3}\right)}{({n_1} {Z_1}+{n_2} ({Z_2}-1)+{n_3}
   (-1+{Z_3})) ({n_1} {Z_1}+{n_2} {Z_2}+{n_3} {Z_3})}. 
 \end{array}
 \label{E1E2Projection}
\end{equation}
By simple but tedious computations we can check that these vectors belongs to the projection plane:
\begin{equation}
 E_{1}\cdot n = 0, \quad E_{2}\cdot n = 0, \quad E_{3} \cdot n=0.
\end{equation}
Practically these vectors can be obtained when adding four distinguishable markers (for example different metal balls) in the phantom in the positions $0$, $e_{1}$, $e_{2}$, and $e_{3}$. 
In addition, we can normalize these vectors 
\begin{equation}
  E_{1n}= \frac{E_{1}}{|E_{1}|}, \quad  E_{2n}= \frac{E_{2}}{|E_{2}|}, \quad E_{3n}= \frac{E_{3}}{|E_{3}|}.
\end{equation}
We can also derive cosine of the angle between $E_{1}$ and $E_{2}$ as a complicated equation
\begin{equation}
\begin{array}{c}
 cos(\angle(E_{1},E_{2})) = E_{1n}\cdot E_{2n} = \\
 -\frac{({n_1} {Z_1}+{n_2} ({Z_2}-1)+{n_3} {Z_3}) \left({n_1}^2
   {Z_1} {Z_2}+{n_1} {Z_3} ({n_3} {Z_2}-{n_2}
   {Z_3})+{n_2} {Z_1} ({n_2} {Z_2}+{n_3} {Z_3})\right)}{\left({Z_1}^2
   \left({n_1}^2+{n_2}^2\right)+2 {n_1} {n_3} {Z_1}
   {Z_3}+{Z_3}^2 \left({n_2}^2+{n_3}^2\right)\right) ({n_1}
   ({Z_1}-1)+{n_2} {Z_2}+{n_3} {Z_3}) } \times \\
   \frac{\sqrt{\frac{L^2 \left({n_1}^2+{n_2}^2+{n_3}^2\right)^2 \left({Z_1}^2
   \left({n_1}^2+{n_2}^2\right)+2 {n_1} {n_3} {Z_1}
   {Z_3}+{Z_3}^2 \left({n_2}^2+{n_3}^2\right)\right)}{({n_1}
   {Z_1}+{n_2} ({Z_2}-1)+{n_3} {Z_3})^2 ({n_1} {Z_1}+{n_2}
   {Z_2}+{n_3} {Z_3})^2}}}{\sqrt{\frac{L^2
   \left({n_1}^2+{n_2}^2+{n_3}^2\right)^2 \left({n_1}^2
   \left({Z_2}^2+{Z_3}^2\right)+({n_2} {Z_2}+{n_3}
   {Z_3})^2\right)}{({n_1} ({Z_1}-1)+{n_2} {Z_2}+{n_3} {Z_3})^2
   ({n_1} {Z_1}+{n_2} {Z_2}+{n_3} {Z_3})^2}}},
\end{array}
\label{Eq:CosE1E2}
\end{equation}
and similarly for angles between other axes. These relations can be used to formulate equations for unknown $Z$ and $Ln$ however we can use additional balls for this purpose  as well. For a ball which center in global coordinates has position given by the vector $b=(b_{1},b_{2},b_{3})$ the projection onto $P$ plane is given by the following formula:
\begin{equation}
\begin{array}{l}
 B = \frac{(-Z+b)\wedge n}{(-Z+b)\cdot n} Ln = \frac{L}{{{b_1} {n_1}+{b_2}
   {n_2}+{b_3} {n_3}-{n_1} {Z_1}-{n_2} {Z_2}-{n_3}
   {Z_3}}}(e_3 ({n_3} (-{b_1} {n_1}-{b_2} {n_2}+{n_1}
   {Z_1}+{n_2} {Z_2})+ \\
   {b_3} \left({n_1}^2+{n_2}^2\right)-{Z_3}
   \left({n_1}^2+{n_2}^2\right))-{b_1} e_2 {n_1} {n_2}+{b_1}
   e_1 {n_2}^2+{b_1} e_1 {n_3}^2+{b_2} e_2 {n_1}^2- \\
   {b_2} e_1 {n_1} {n_2}+{b_2} e_2 {n_3}^2-{b_3} e_1 {n_1}{n_3}-{b_3} e_2 {n_2} {n_3}-e_2 {n_1}^2 {Z_2}+ \\   
   e_2 {n_1} {n_2} {Z_1}+e_1 {n_1} {n_2} {Z_2}+e_1 {n_1} {n_3}
   {Z_3}-e_1 {n_2}^2 {Z_1}+e_2 {n_2} {n3} {Z_3}-e_1 {n_3}^2 {Z_1}-e_2 {n_3}^2 {Z_2}).
\end{array}
\label{Eq.BallProjection}
\end{equation}
Finally, the projection of the beam $a$ onto $A$ is given by
\begin{equation}
 \begin{array}{l}
  A=\frac{a\wedge n}{a\cdot n} Ln = \frac{L}{{a_1} {n_1}+{a_2}{n_2}+{a_3}{n_3}}(({a_1} (-e_2 {n_1} {n_2}-e_3 {n_1} {n_3}+e_1 {n_2}^2+e_1 {n_3}^2)+ \\
  {a_2} (e_2 {n_1}^2-e_1 {n_1} {n_2}+{n_3} (e_2 {n_3}-e_3 {n_2}))+{a_3} e_3 ({n_1}^2+{n_2}^2)-{a_3} {n_3} (e_1 {n_1}+e_2 {n_2})).
 \end{array}
 \label{Eq.BeamProjection}
\end{equation}

We have now all formulas to fully specify an algorithms for calibration of a radiotherapeutic device.

\section{Alignment algorithm}

\subsection{Mechanical alignment}
\begin{figure}
\centering
 \includegraphics[width = 0.5\textwidth]{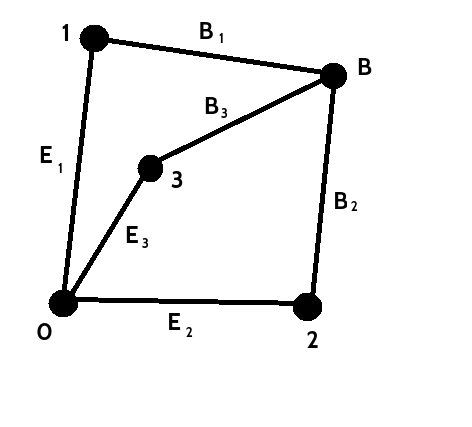}
 \caption{The picture on projection plane and the distances to be measured.}
 \label{Fig.MechanicalAlignemnt}
\end{figure}

In the first stage for a fixed position  $\alpha$ of a gantry we have to calculate $6$ parameters - $Ln_{\alpha}$ and $Z_{\alpha}$, therefore we have to specify six equations. They can be proposed in different ways depending on the accuracy of measurement on imaging plane $P$. Here are some of the examples
\begin{itemize}
 \item {Angles between projected axes, namely: $cos(\angle(E_{1},E_{2}))$, and $cos(\angle(E_{2},E_{3}))$ (the third cosine is dependent from the given ones).}
 \item {We can provide one additional ball in the phantom at the position $b$ and we have $6$ additional equations - 3 for $\{|E_{i}|\}_{i=1}^{3}$ and  $|B_{i}| = \left|\frac{(-Z+b_{i})\wedge n}{(-Z+b_{i})\cdot n} Ln -  \frac{(-Z+e_{i})\wedge n}{(-Z+e_{i})\cdot n} Ln \right|$ as it is presented in Fig. \ref{Fig.MechanicalAlignemnt}.}
\end{itemize}
We can use any combination of these equations. In practice one can also add additional balls in order to obtain more equations and then apply optimization technique to recover $L$,  $n_{\alpha}$ and $Z_{\alpha}$.

\subsection{Radiation alignment}
\begin{figure}
\centering
 \includegraphics[width = 0.5\textwidth]{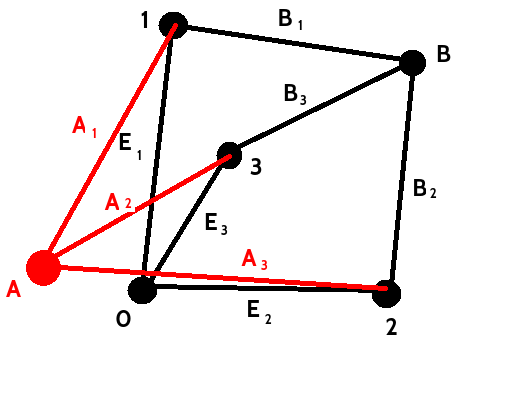}
 \caption{The picture on projection plane and the distances to be measured.}
 \label{Fig.RadiationAlignemnt}
\end{figure}

According to the proposed procedure a point $A$ at the intersection of the central axis of the radiation field and the detector plane must be selected first. This point can be for example in the center of a rectangular projection of collimator jaws or a projection of a selected point within a cancer which should receive the treatment. Knowing $n_{\alpha}$ and $Z_{\alpha}$ for fixed $\alpha$ from the previous step, the position of $A$ with respect to the projections of fiducials $1$, $2$, $3$, defining the phantom-related frame of reference (see Fig. \ref{Fig.GeneralSetupCalibrationSource}) have to be measured, i.e., 
\begin{equation}
 |A_{i}|= \left|A- \frac{(-Z+e_{i})\wedge n}{(-Z+e_{i})\cdot n} Ln \right|, \quad i\in\{1,2,3\},
\end{equation}
as presented in Fig. \ref{Fig.RadiationAlignemnt}.

This is a minimal number of equations that allows to restore tree coordinates of $a_{\alpha}=(a_{1,\alpha}, a_{2,\alpha},a_{3,\alpha})$. However, for error reduction more distances should be measured.

Next, knowing $a_{\alpha}$ and $Z_{\alpha}$ one can derive required correction of the phantom origin position (\ref{Eq.RadiationSourceCorrection}) in order to get the phantom origin aligned with the central axis of the radiation beam. For this setup we get a separate shift vector $\Delta Z_{\alpha}$ for each $\alpha$ using (\ref{Eq.RadiationSourceCorrection}).

An alternative approach is to find some average correction $\Delta\overline{Z}$, which is, on average, good for all angles $\{\alpha_{i}\}_{i=1}^{N}$, where $N$ is the number of calibration positions. For example one can minimize the following cost function $F(\overline{\Delta Z})$:

\begin{equation}
 F(\overline{\Delta Z}) = \sum_{\alpha} \frac{1}{a_{\alpha}^{2}} |(\overline{\Delta Z} + Z_{\alpha} )  \times a_{\alpha}|^{2},  
 \label{Eq.averageSetup}
\end{equation}
where $\times$ is a cross product, a dual to $\wedge$. This form of the cost function is evident when combining (\ref{Eg.BeamAlignment}) with dualization (\ref{Eq.dualization}).

\section{Summary}
In the present paper a general setup for determination of imperfect geometry of radiotherapeutic devices has been proposed based on geometric algebra framework. In reality accuracy of the geometric information provided by a radiotherapeutic system is limited by the geometric stability of a therapeutic system system - flexing or sagging of this system under its own weight during gantry rotation results in non-ideal source and detector plane positions during gantry rotations. For this reason a notion of isocentre of an ideal therapeutic device must be relaxed in real laboratory settings - instead for each nominal angular position of a gantry we have a separate radiation axis. Axes determined for different angular position of a gantry do not intersect in a single point of an ideal isocenter. Clearly, the most generic calibration should thus consists of determinating for each angular position of a gantry how an origin of a laboratory frame of reference must be shifted to be placed along a radiation axis for this angular position.

In this method two methods of a calibration so defined were presented differing in the assumed purpose. Independently on the chosen method the calibration concepts are expressed in an extremely simple and natural way using tool from geometric algebra framework.

The proposed calibration methods can be used in practice in the following way. Assume that the phantom has been positioned on a treatment table according to the room laser pointers so that the phantom center is pointed by the lasers, that is the room lasers define the origin of the laboratory frame of reference.  Then, to account for the imperfect geometry of a therapeutic device the introduced methodology tells us how the phantom must be shifted for a given angular position $\alpha$ of a gantry to have its center ideally in the central axis of the radiation field. The corrections calcuated according to the developed methodology can be applied separately for each $\alpha$ if we use (\ref{Eq.RadiationSourceCorrection}) to get the most accurate performance or it can be used only once, according to (\ref{Eq.averageSetup}) to get a treatment result which is on average optimal. 
Then, if the calculated shift exceeds some acceptance level, laser pointers or other components of an EBT device should be recalibrated based on the results of correction assessment.

\section*{Acknowledgments}
This research was supported by  the grant POIR.04.01.04-00-0014/16 of The National Centre of Research and Development. RK would like to thanks prof. Julian \L{}awrynowicz for introducing him the beautiful world of Clifford algebras and their applications.

\appendix

\section{Mathematica calculations using Cartan package}
\label{Appendix:MathematicaPackage}
All functions useful in our calculations from Cartan package \cite{GAMathematica} is summarized in Tab. \ref{Tab:CliffordMathematica}.
\begin{table}
\centering
\begin{tabular}{ | l | l |}
    \hline
    e[1],e[2],e[3] & $e_{1},e_{2},e_{3}$ \\ \hline
     GeometricProduct[a,b] & $ab$ \\ \hline
     InnerProduct[a,b] & $ a \cdot b$ \\ \hline
     OuterProduct[a,b] & $ a \wedge b$ \\ \hline
     Turn[a]   & $a^{\dagger}$ \\ \hline
\end{tabular} 
\caption{Selected basic functions from Cartan package \cite{GAMathematica}.}
\label{Tab:CliffordMathematica}
\end{table}
The package can be initialized using (see \cite{GAMathematica})
\begin{verbatim}
 << clifford.m
\end{verbatim}
In order to simplify geometric operations one can define a few new functions. Projection of a vector $a$ fixed at the source point onto plane given by the vector $n$ that connects source and the plane and perpendicular to the plane is realized by the function
\begin{verbatim}
 ProjectOntoPlane[a_, n_] := 
 GeometricProduct[OuterProduct[a, n], n]/InnerProduct[n, a]
\end{verbatim}

Rotations of a vector $a$ in the plane given by the blade $b$ by the angle $\alpha$ can be defined as follows
\begin{verbatim}
 RotateVector[v_, b_, \[Alpha]_] := 
 Module[{R}, R[\[Beta]_] := Cos[\[Beta]] - b*Sin[\[Beta]]; 
  Return[FullSimplify[
    GeometricProduct[R[\[Alpha]/2], v, Turn[R[\[Alpha]/2]]]]]]
\end{verbatim}

Norm of a vector can be realized by
\begin{verbatim}
 VectorNorm[a_] := Sqrt[InnerProduct[a, a]]
\end{verbatim}

In these terms all our calculations presented so far can be recovered a few lines of code:
\begin{verbatim}
 n = n1*e[1] + n2*e[2] + n3*e[3]
 Z = Z1*e[1] + Z2*e[2] + Z3*e[3]
 (*project center:*)
 Zp = Simplify[ProjectOntoPlane[-Z, L*n]]
 (*project base vectors:*)
 E1 = FullSimplify[ProjectOntoPlane[-Z + e[1], L*n] - Zp]
 E2 = FullSimplify[ProjectOntoPlane[-Z + e[2], L*n] - Zp]
 (*trivial check if projected vectors lie in the plane:*)
 InnerProduct[E1, n]
 InnerProduct[E2, n]
 (*rotate base vectors and then project:*)
 E1r = FullSimplify[ ProjectOntoPlane[-Z + RotateVector[e[1], e[2] e[3], 
 \[Alpha]], L*n] - Zp]
 E2r = FullSimplify[ ProjectOntoPlane[-Z + RotateVector[e[2], e[2] e[3], 
 \[Alpha]], L*n] - Zp]
 (* cosine of the angle between projected vetors - time consuming!:*)
 CosE1rE2r = FullSimplify[InnerProduct[ E1r/VectorNorm[E1r],E2r/VectorNorm[E2r]]]
 (*vertex definition:*)
 v = Table[0, {i, 1, 8}]
 v[[1]] = +a*e[1] + a*e[2] + a*e[3]
 v[[2]] = +a*e[1] + a*e[2] - a*e[3]
 v[[3]] = +a*e[1] - a*e[2] + a*e[3]
 v[[4]] = +a*e[1] - a*e[2] - a*e[3]
 v[[5]] = -a*e[1] + a*e[2] + a*e[3]
 v[[6]] = -a*e[1] + a*e[2] - a*e[3]
 v[[7]] = -a*e[1] - a*e[2] + a*e[3]
 v[[8]] = -a*e[1] - a*e[2] - a*e[3]
 (*projection using parallel computing:*)
 vP = ParallelMap[ FullSimplify[ ProjectOntoPlane[-Z + RotateVector[#, 
 e[2] e[3],  \[Alpha]], L*n]] &, v]
 (*Projection of projected corners on the E1r, E2r base:*)
 MapThread[
 Print["vector = ", #1, "; E1r*v = ", 
   FullSimplify[InnerProduct[#2, E1r]],  "; E2r*v = ", 
   FullSimplify[InnerProduct[#2, E2r]]] &, {v, vP}]
\end{verbatim}




\end{document}